\documentclass[reprint]{revtex4-1}
\usepackage{amsmath,amssymb,bm}
\usepackage{graphicx}
\usepackage{epstopdf}
\usepackage{latexsym}
\usepackage{subfigure}
\usepackage{color}

\begin{document}

\title{Quantum criticality in SU(3) and SU(4) JQ anti-ferromagnets}
\author{Ribhu K. Kaul}
\affiliation{Department of Physics \& Astronomy, University of Kentucky, Lexington, KY-40506-0055}
\begin{abstract}
We study the quantum phase transition out of the N\'eel state in SU(3) and SU(4) generalizations of the Heisenberg anti-ferromagnet with a sign problem free four spin coupling (so-called JQ model), by extensive quantum Monte Carlo simulations. We present evidence that the SU(3) and SU(4) order parameters and the SU(3) and SU(4) stiffness' go to zero continuously without any evidence for a first order transition. However, we find considerable deviations from simple scaling laws for the stiffness even in the largest system sizes studied. We interpret these as arising from multiplicative scaling terms in these quantities which affect the {\em leading} behavior, {\em i.e.}, they will persist in the thermodynamic limit unlike the conventional additive corrections from irrelevant operators. We conjecture that these multiplicative terms arise from {\em dangerously} irrelevant operators whose contributions to the quantities of interest are non-analytic.
\end{abstract}
\maketitle

\section{Introduction}

The theory of quantum phase transitions~\cite{sachdev1999:qpt} has become a centerpiece in the research of quantum physics of strongly coupled condensed matter.
It has found applications in many branches of condensed matter physics, examples include: heavy fermion systems~\cite{si2010:heavy}, high temperature superconductivity~\cite{kaul2008:acl}, quantum Hall effect~\cite{sondhi1997:qpt}, metal-insulator transition, ultra-cold atomic gases~\cite{nikolic2007:bcsbec} and frustrated magnets~\cite{coldea2010:E8,lee2010:columbite}.
Aside from its many applications to natural systems, the study of quantum phase transitions has also turned into a full fledged theoretical endeavor in its own right, which although very mature in certain aspects~\cite{sachdev1999:qpt}, is still
in a stage of infancy when one considers the large number of different physical situations modern condensed matter systems allow for.
Indeed, most of the well understood quantum critical points (continuous quantum phase transitions) can be simply related to a classical universality class in one higher dimension. 
%The paradigm for such a mapping is the transverse field Ising chain and its mapping to the well understood two dimensional Ising field theory. 
On the other hand, one of the most exciting directions in the field is the study of quantum critical points which lead to new universalities that are not natural to think of in the
classical context. Such new criticality may arise, for instance, from essentially quantum phenomena such as the presence of low-energy fermions or complex Berry phases that have no natural classical analogue.

An interesting theoretical proposal for a quantum critical point that does not have a naive classical analogue was put forward a few years ago~\cite{senthil2004:science,senthil2004:deconf_long}.
The physical system is a two-dimensional quantum anti-ferromagnet on a square lattice with S=1/2 spins which transitions into a paramagnet with a broken 
translational symmetry, a valence bond solid (VBS). In a set of compelling arguments, it was shown that a naive application of the classical theory which forbids a continuous transition could be invalidated by the presence
of Berry phase terms, giving rise to a quantum critical point in a novel universality class. It is widely accepted, however, that a full non-perturbative understanding of the occurrence and properties of the new quantum critical point
are beyond analytic reasoning and can only be established by unbiased numerical simulations. 
In a pioneering piece of work~\cite{sandvik2007:deconf},  it was shown that these questions could be addressed 
in a sign problem free microscopic JQ model using quantum Monte Carlo techniques. Since then a number of works have studied this SU(2) symmetric model~\cite{melko2008:jq,jiang2008:first,kaul2008:jq,sandvik2010:log}. While all workers agree that
the JQ model harbors a N\'eel state on one side and  a VBS state on the other side, a clear picture of the
the nature and precise scaling at the quantum phase transition still does not exist. The most comprehensive numerical
study of the SU(2) JQ model however provides strong evidence in favor of a continuous transition~\cite{sandvik2010:log}, although with corrections to the naive scaling hypothesis that affect the leading behavior. 
Corrections to naive scaling also appear in the study of the response of the JQ quantum critical point to single impurities~\cite{banerjee2010:log}. 
Another study~\cite{harada2006:deconf} for an unconventional quantum critical point in the spatially anisotropic bilinear bi-quadratic model has also
found evidence for a continuous quantum critical point, though this model is less well studied. The only theory for this transition is also beyond
a naive classical order parameter theory~\cite{grover2007:deconf}.
Excluding the JQ model and the anisotropic bilinear biqudratic $S=1$ model, there is no known
candidate for a quantum critical point in a numerically accessible two dimensional microscopic quantum Hamiltonian that does not map simply onto a higher dimensional classical field theory (the most well studied example of a quantum critical point with a higher dimensional classical mapping is the bilayer anti-ferromagnet, the mapping has been studied thoroughly~\cite{wang2006:bilayer}).
Sorting out these issues is a very demanding one numerically, but is extremely important for the field of quantum criticality.

In this paper we present the results of extensive QMC simulations on a larger $N$ extension of the original SU($N=2$) symmetric JQ model.
%We focus on the criticality of the SU($N$) order parameter and the scaling of the spin stiffness and susceptibility, by generalizing the JQ model 
%to SU(3) and SU(4) symmetry. 
The first study~\cite{lou2009:sun} of these models reported $T=0$ QMC results using
the valence bond basis method for the SU(3) and SU(4) models and found results consistent with a continuous transition with conventional scaling
for quantities associated with the VBS and N\'eel order parameters.
 In this work we approach this model with the powerful finite-T stochastic series expansion (SSE)
method~\cite{syljuasen2002:dirloop} which allows access to finite-temperature, larger volumes and most importantly for our purposes here, gives access to the stiffness and susceptibility. We find that both the SU(3) and SU(4) models have continuous phase transitions with no evidence for first order behavior even on the largest volumes studied in this work, in agreement with the past work on smaller sizes.
Our main new finding is the existence of  {\em multiplicative} power law terms in the scaling of the stiffness and the susceptibility that affect {\em the leading} scaling behavior for the SU(3) and SU(4) models. We conjecture that the breakdown of conventional scaling
behavior is due to the presence of a {\em dangerously irrelevant operator}, which although formally irrelevant can affect leading scaling behavior in physical quantities if the latter depend non-analytically on the dangerously irrelevant coupling.
The layout of the paper is as follows: In Sec.~\ref{sec:model}, we describe in detail the model and the observables we study in this paper. In Sec.~\ref{sec:binder} and~\ref{sec:stiff} , we discuss the scaling properties of the Binder ratio and the spin stiffness respectively. Finally, we conclude in Sec.~\ref{sec:discuss} with a discussion of our main results.

Larger-$N$ extensions of the SU(2) physics may have applications to
cold atom systems~\cite{gorshkov2010:sun}, but are important in their own right as they allow one to approach the analytically well understood large$-N$ limit~\cite{senthil2004:deconf_long}. Other numerical work has studied the phase transition between N\'eel-VBS by varying the $N$ in SU($N$) continuously~\cite{beach2009:sun}, without the introduction of a four spin $Q$ term.

\section{Model}
\label{sec:model}

\begin{figure}[t]
 \includegraphics[width=3.5in]{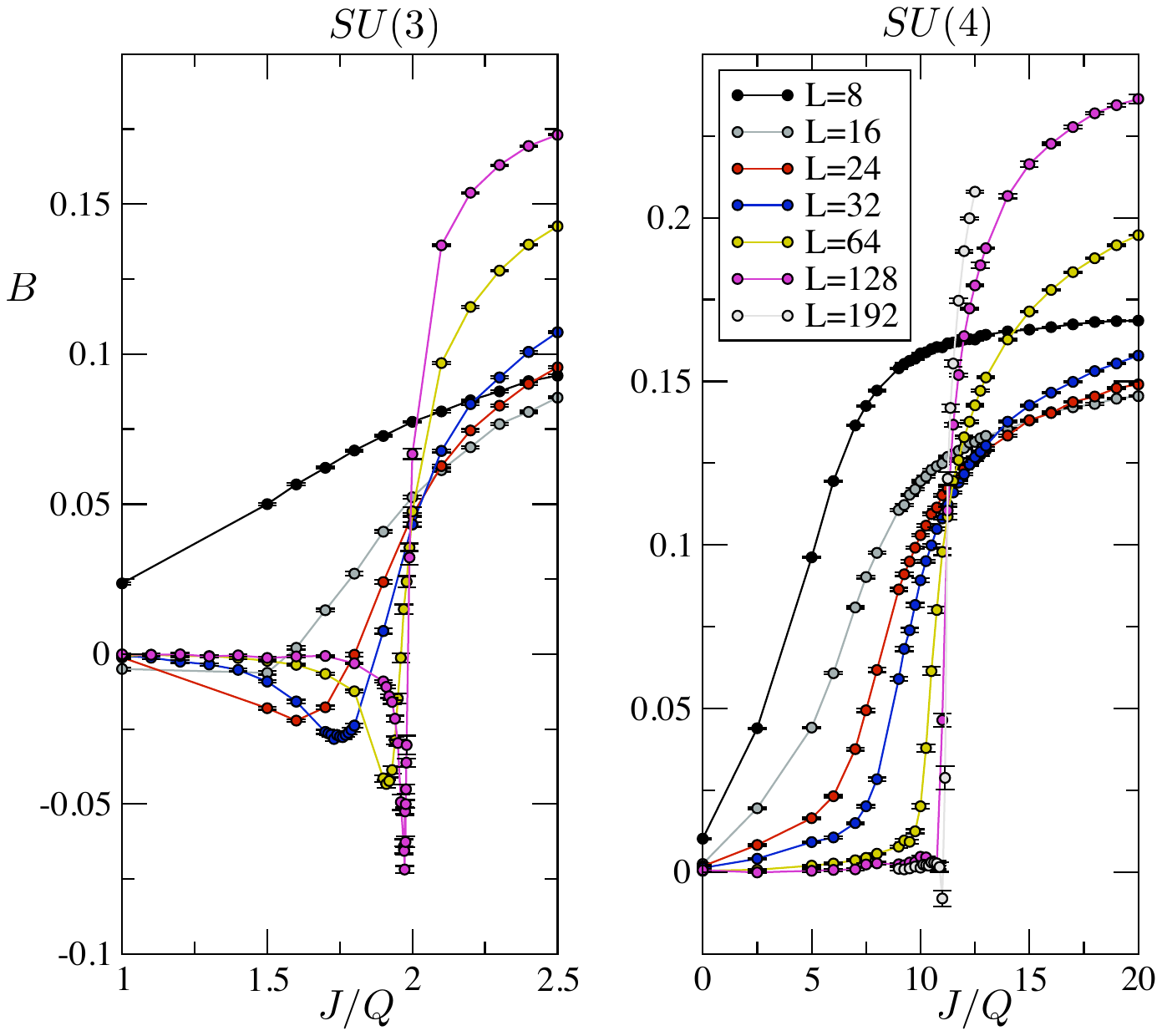}
   \caption{\label{fig:binder_bb} Binder ratio of the JQ model for SU(3) and SU(4). Note particular the fact that the Binder ratio becomes negative.
For the SU(3) case it is clear that the minimum does not saturate with system size. For the SU(4) system the negativity appears only at very large systems sizes and hence we cannot make a conclusion on the behavior in the thermodynamic limit. As explained in the text in Sec.~\ref{sec:binder} the divergence is far too slow to indicate a first order transition.}
\end{figure}

\begin{figure}[t]
 \includegraphics[width=3.5in]{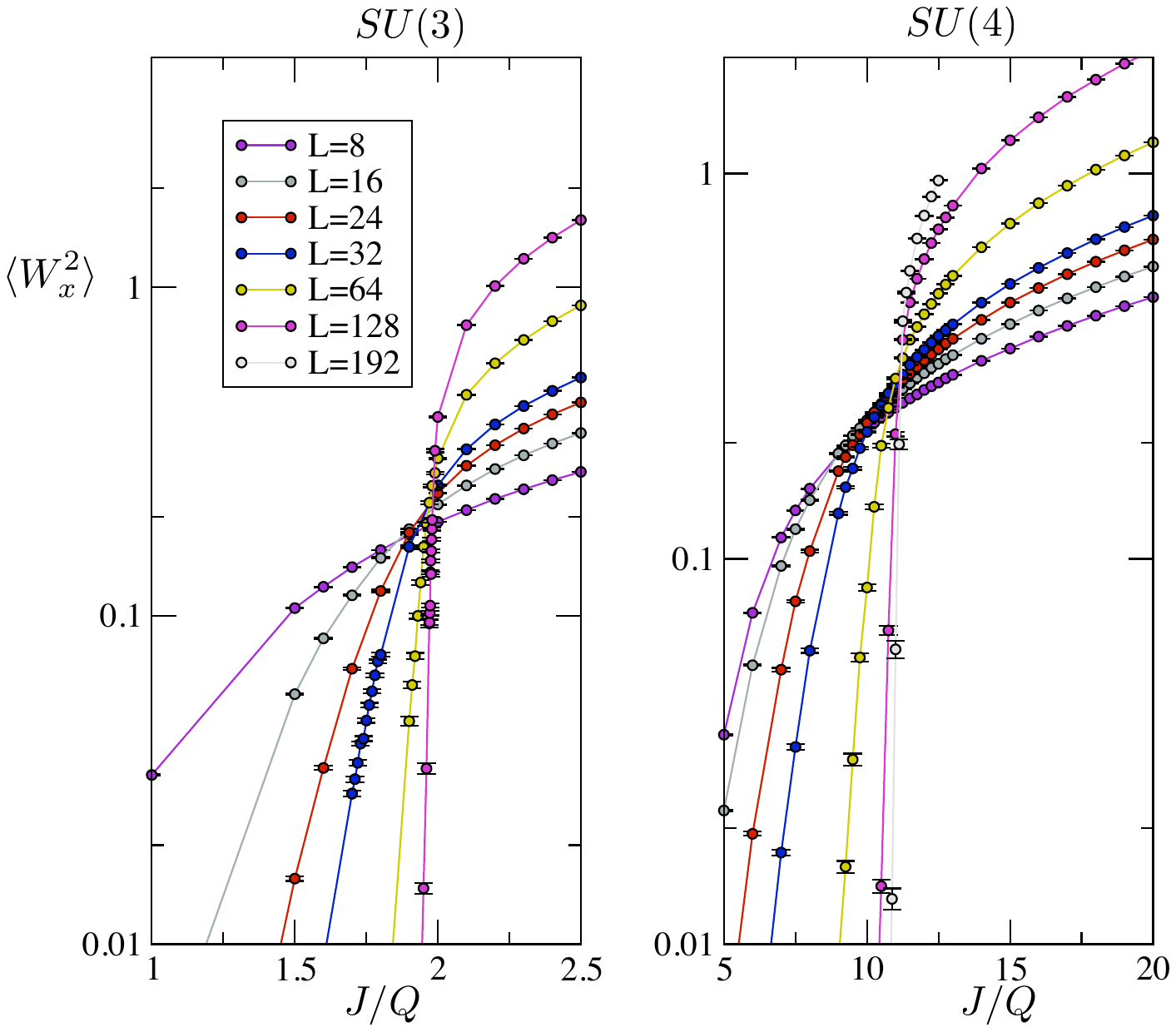}
   \caption{\label{fig:chirho_bb} $\langle W_x^2\rangle$ data for the same set of runs as Fig.~\ref{fig:binder_bb}. The y-axis is on a log scale. There is clear evidence for a crossing in the data. The crossing is analyzed in detail in Sec.~\ref{sec:stiff}}
\end{figure}

We define our model on a square lattice, which can be conveniently separated into an A and B sub-lattice. We define $N$
flavors of fermions, $f_\alpha$ identically on each site ($1\leq\alpha\leq N$). An SU($N$) rotation will rotate
among these $N$ flavors with the following convention:
\begin{equation}
f_\alpha \rightarrow f^\prime_\beta = U^*_{\alpha\beta}f_\beta
\end{equation}
where the $^*$ implies complex conjugation. Now to pick a spin
representation in terms of the $f_\alpha$ fermions, we fix the number of fermions on each site, $f^\dagger_\alpha
f_\alpha =1 $ on the A sub-lattice and $f^\dagger_\beta f_\beta =
(N-1)$ on the B sub-lattice. This fixes the size of the local Hilbert space to be simply $N$ on each site, on both the
A and on the B sub-lattice. We define these $N$ states with the following sign conventions 
and indicate their transformations under SU($N$) rotation,
\begin{eqnarray}
|\alpha \rangle_A &=& f^\dagger_\alpha |0 \rangle~~~ (|\alpha\rangle_A
\rightarrow U_{\alpha\beta}|\beta\rangle_A)\\
|\alpha \rangle_B &=& f_\alpha|F\rangle ~~~ (|\alpha\rangle_B
\rightarrow U^*_{\alpha\beta}|\beta\rangle_B)
\end{eqnarray}
where $|0\rangle$ denotes the absence of any fermions and $|F\rangle$
is a fully filled site. The transformation properties imply that the $N$, A sub-lattice states transform in
the fundamental representation of SU($N$) and the $N$, B sub-lattice states transform in the conjugate of the fundamental representation.
The transformations also imply that the state
$\sum_\alpha |\alpha\rangle_A|\alpha\rangle_B$ transforms as an SU($N$) singlet. This is the main reason we have chosen different representations on the A and B sub-lattices, {\em i.e.} to allow the formation of a two-site singlet, following previous work on anti-ferromagnets~\cite{marston1988:sun,read1990:vbs,harada2003:sun}.

To construct a spin model, we write down SU($N$) invariant four fermion interactions that
maintain the number of fermions on each site; Explicitly, there are two such
terms $f^\dagger_{i\alpha} f_{i\alpha}f^\dagger_{j\beta} f_{j\beta}$
and $f^\dagger_{i\alpha} f_{i\beta} f_{j\alpha}f^\dagger_{j\beta}$,
the first one is just the identity operator in the projected space,
while the second term which we call, $P_{ij}$ is a projector onto an $ij$ singlet ($i$ is on A sub-lattice and $j$ is on B sub-lattice). 
Note that the matrix elements of $P_{ij}$ are very simple,
\begin{equation}
\langle \alpha_1 \beta_1 | P_{ij}|\alpha_2 \beta_2\rangle = \delta_{\alpha_1\beta_1}\delta_{\alpha_2\beta_2}
\end{equation}
Since they are always positive when non-zero, we can use these operators to construct sign problem free models. 
In particular the model defined by $H_{ij}=-\frac{P_{ij}}{2}$ is the familiar Heisenberg model up to a constant, 
 after identifying $\alpha=1$ with up(down) and $\alpha=2$ with down(up) on the A(B) sub-lattice. We define the SU($N$) 
``JQ'' model with the following conventions.
\begin{equation}
H_{JQ} = -\frac{J}{N}\sum_{ij}P_{ij}-\frac{Q}{N^2}\sum_{ijkl}P_{ij}P_{kl}
\end{equation}
where the first sum is taken over all nearest neighbor bonds on the square lattice and the second term is taken over all elementary
plaquettes of the square lattice with $ij$ and $kl$ being nearest neighbor bonds. 
Note that so defined, $H_{JQ}$ explicitly has no sign problem: all off-diagonal matrix elements are explicitly negative.

We now turn to the observables of interest in our study here.
There are $N^2-1$ traceless Hermitian matrices, $X^a_{\alpha\beta}$ which generate the
SU($N$) algebra. Of these, $N-1$ can be chosen diagonal (so-called
Cartan subalgebra). We choose them
so that they satisfy: ${\rm Tr}[X^aX^b]=\delta_{ab}/2$. We can then
work out formulas for the ``uniform magnetization'' and ``staggered
magnetization'' for the Cartan generators in terms of the operator $n_\alpha$ which measures
which of the $|\alpha\rangle$ states is occupied on each site
($n_\alpha=f^\dagger_\alpha f_\alpha$ on the A sub-lattice and
$n_\alpha=1-f^\dagger_\alpha f_\alpha$ on the B sub-lattice). 
\begin{eqnarray}
M^a_u &=& \sum_{r\in A}X^a_{\alpha\alpha}n_\alpha-\sum_{r\in
  B}X^a_{\alpha\alpha}n_\alpha\\
M^a_s &=& \sum_{r\in A}X^a_{\alpha\alpha}n_\alpha+\sum_{r\in
  B}X^a_{\alpha\alpha}n_\alpha
\end{eqnarray}
Note that the sign in the middle is opposite from what you would have expected for
the familiar
$S=1/2$ case, because of the way our Hilbert space is defined
 on the A and B sub-lattice. The specific choices for the Cartan generators is detailed in Appendix~\ref{sec:Cartan}. 
An outline of the algorithm and method is given in Appendix~\ref{sec:method}.

\section{Binder Ratio}
\label{sec:binder}

\begin{figure}[t]
 \includegraphics[width=3.5in]{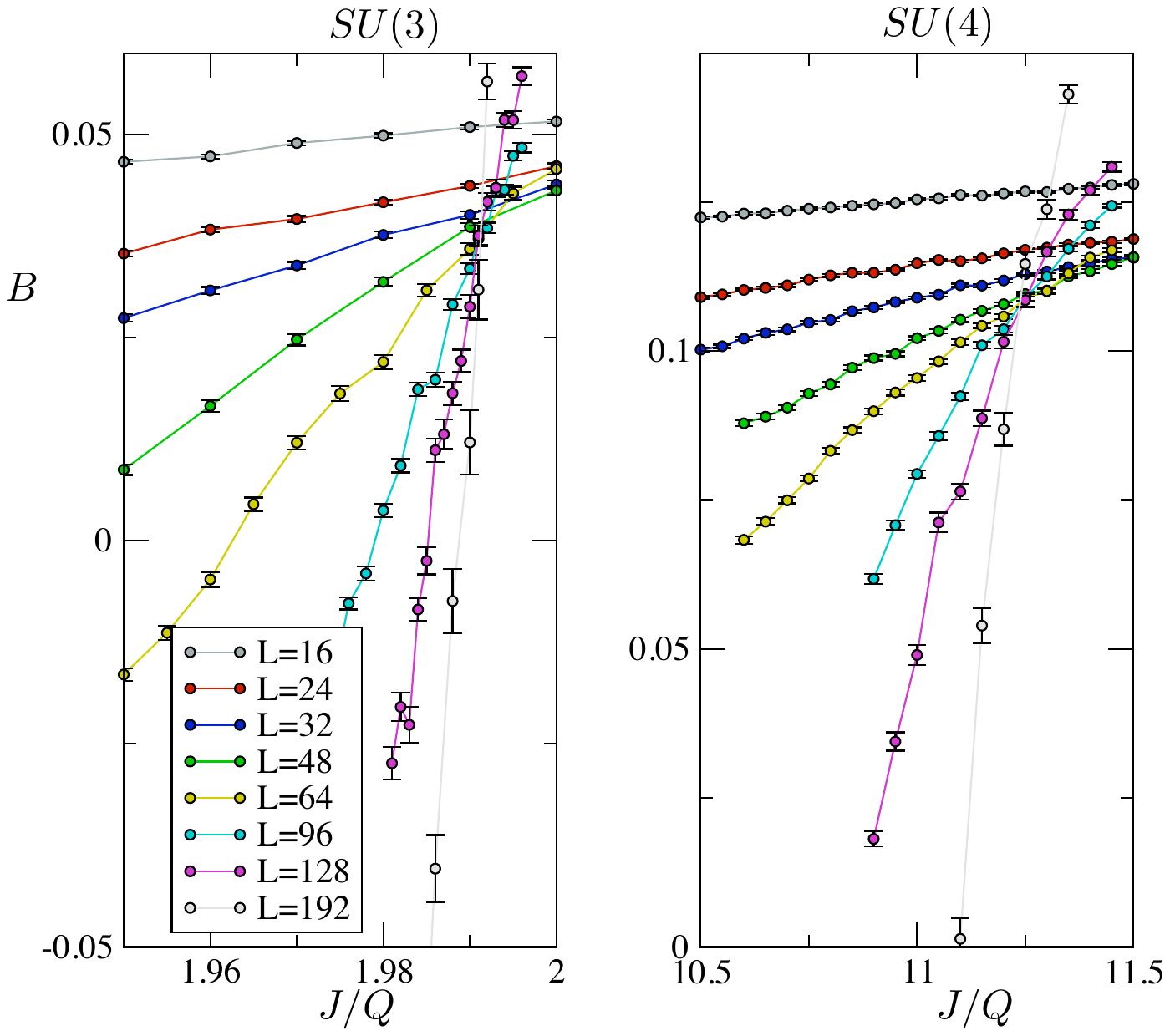}
   \caption{\label{fig:binder} A zoom-in of the crossing of the binder ratio, $B$, for SU(3) and SU(4) symmetric JQ models. Error bars were determined by bootstrapping the data. The solid lines are polynomial fits on the data and are plotted to guide the eye. The crossing point converges well for the larger system sizes. For $L=64,96,128,192$ the data crosses nicely within the estimated error bars, allowing very accurate brackets for the quantum critical points. 
$(J/Q)^3_c=(1.9905,1.9920)$ and $(J/Q)^4_c=(11.235,11.255)$.}
\end{figure}

\begin{figure}[t]
 \includegraphics[width=3.5in]{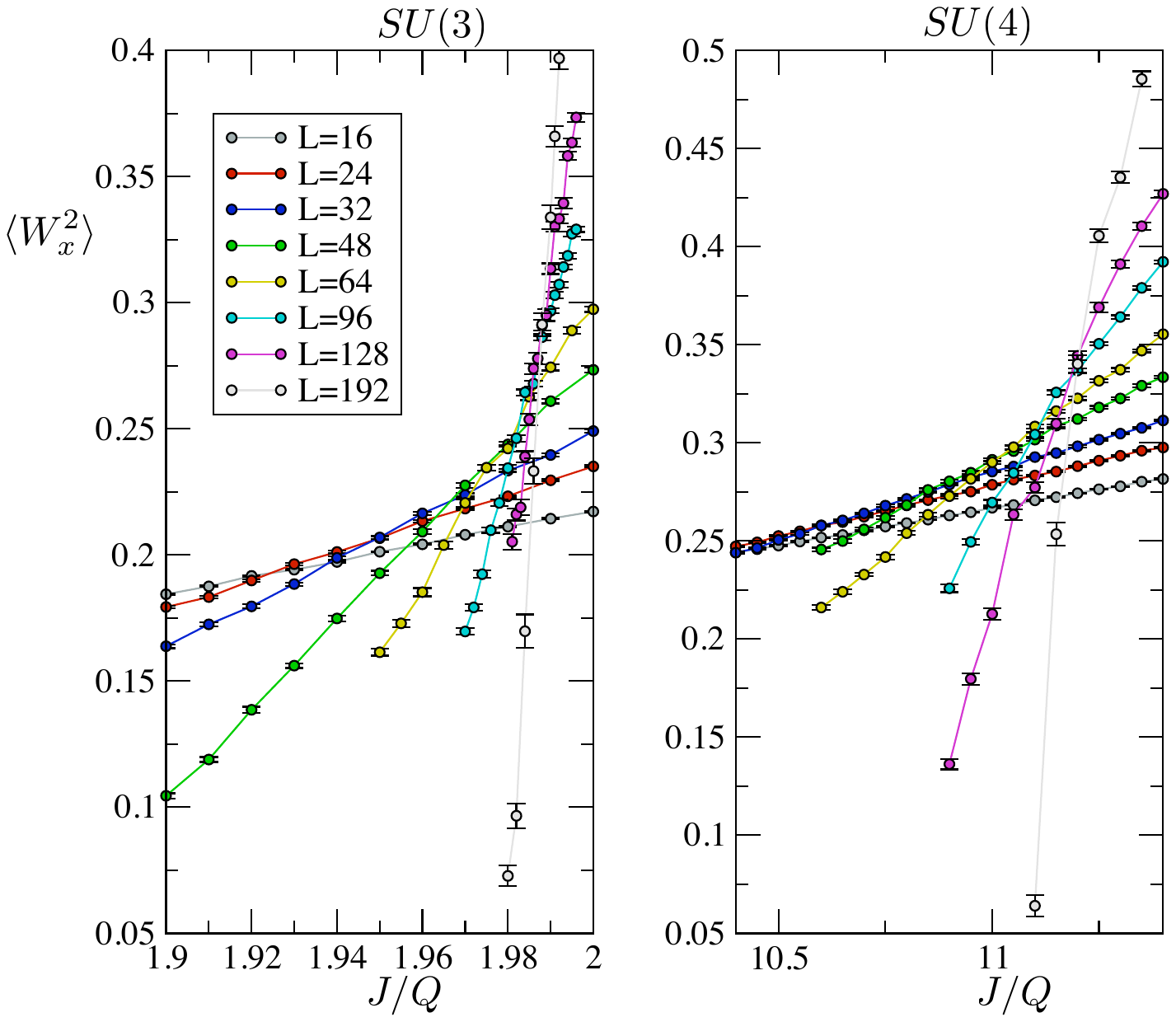}
   \caption{\label{fig:chirho_bare}A zoom-in of the fluctuations of the spatial winding number (stiffness) for SU(3) and SU(4) symmetric JQ models. 
     This is the same set of simulations as for the Binder ratio data shown in Fig.~\ref{fig:binder} (this is a somewhat wider view to accommodate drift of crossing; compare $x-$axes). Clearly there are significant deviations from the simple
scaling behavior expected for a quantity of scaling dimension zero.
}
\end{figure}

The first quantity which we study is the so-called Binder ratio. It is defined as,
\begin{equation}
B = 1-\frac{\langle M_s^4\rangle}{3\langle M_s^2\rangle^2}
\end{equation}
In addition to the Monte-Carlo averaging, the $\langle \dots \rangle$ imply averaging over the $N-1$ Cartan generators.
It is well known that the Binder ratio has a scaling dimension of zero and hence if we assume a standard scaling hypothesis, we can write down the following simple scaling form close to a quantum critical point,
\begin{equation}
B= \mathbb{B}(\frac{L^zT}{c},gL^{1/\nu})
\end{equation}
where $z$ is the (universal) dynamic critical exponent which connects the scaling dimensions of space and time at a quantum critical point, $c$ is a (non-universal) velocity that connects the engineering dimensions of space and time. These two quantities ensure that $L^zT/c$ has neither a scaling nor an engineering dimension. $g$ is an (engineering) dimensionless measure of the deviation from the critical point ({\em e.g.} $J/Q- (J/Q)_c$) and $\nu$ is a (universal) critical exponent that ensures the parameter $gL^{1/\nu}$ has neither an engineering or scaling dimension. The scaling form implies that if we fix the parameter $LT$ (see Appendix~\ref{sec:lt}), we assume everywhere in this work that $z=1$, that $B$ should 
become volume independent when $g=0$ in the scaling limit ({\em i.e.} for large $L$). The simplest way to find whether this is true is to
plot $B(g)$ for different values of $L$ and look for a crossing of the various curves.  Interesting observations emerge on a detailed study of the quantity $B$: First, $B$ turns negative just before the critical point for both SU(3) and SU(4), Fig.~\ref{fig:binder_bb}. This negative $B$ behavior is clear for the SU(3) model. For the SU(4) this tendency appears only at the largest volume ($L=192$) whereas for SU(3) it already appears at $L\geq 16$, we can hence make better extrapolations on this particular behavior for the SU(3) case: It is clear for SU(3) that the negative minimum value diverges with increasing system size. A quick look at the data indicates that it diverges sub-linearly (we were unable to fit a precise power law form to the Binder minimum because of the large errors in determining the minimum). For a first order transition in a classical model, the minimum diverges as the volume~\cite{vollmayr1993:first}; Analogously, for a quantum first order transition it should diverge as $L^2\beta$ for a zero frequency Binder ratio or as $L^2$ for an equal-time Binder ratio; Our calculations are for an equal time Binder ratio and hence a first order transition should show an $L^2$ divergence. First order $L^2$ diverging behavior has been observed for another JQ model which has been shown to have a first order transition~\cite{sen2010:first}. The fact that our Binder ratio diverges much slower than $L^2$ (it diverges sub-linearly) is hence strong proof that the negative minimum does not indicate a first order transition. The second interesting observation is the presence of a nice crossing that quickly converges to a common point as the volume is increased. This is one of the classic signatures of a continuous transition. We have collected data on system sizes ranging from $L=16-192$. While the data is accurate enough to detect
corrections to scaling for 
the smaller sizes
the crossing converges very well as the system size is increased and the four biggest system sizes cross nicely within our estimated error bars. 
From this crossing we have an accurate estimate of the value of the critical coupling (see caption of Fig.~\ref{fig:binder}), which is 
 consistent with the values quoted in Ref.~\onlinecite{lou2009:sun} [thanks to the larger volumes simulated here, our brackets are about a factor of 2 times more accurate for 
both SU(3) and SU(4)].

\section{Stiffness}
\label{sec:stiff}

Configurations of our SU($N$) model in the SSE method (see Appendix~\ref{sec:method}) 
can be thought of as a set of $N$ colors of non-intersecting closed loops. We can measure
the winding number of each of the $N$ colors of loops both in  space and in time and associate a 
spatial and temporal current for each configuration, with each of the $N-1$ diagonal generators.
\begin{eqnarray}
j_x^a&=&X^a_{\alpha\alpha}W^{x}_{\alpha}\\
j_\tau^a&=&X^a_{\alpha\alpha}W^{\tau}_{\alpha}
\end{eqnarray}
The fluctuations of these quantities averaged over MC sampling and the $N-1$ diagonal generators
$\langle W_x^2\rangle \equiv \langle \left( j^a_x\right)^2\rangle$ and $\langle W_\tau^2\rangle \equiv \langle \left(j^a_\tau\right)^2\rangle$,
are of great interest, since they too are expected to have no scaling dimension just like the Binder ratio, $B$, discussed earlier, because they
are susceptibilities of the generators of a conservation law (SU($N$) symmetry).
Hence, for the same reasons discussed in the case of $B$, they are expected to have a crossing point. Before turning to a numerical study of their crossing, we note that by linear response theory, the fluctuations in spatial winding number are simply related to the familiar spin stiffness ($\rho_s=\langle W_x^2\rangle/\beta$) and the fluctuation in the temporal winding number are related to the spin susceptibility ($\chi_u=\langle W_\tau^2\rangle\beta/L^2$), we shall refer to these interchangeably.

\begin{figure}[t]
  \includegraphics[width=3.5in]{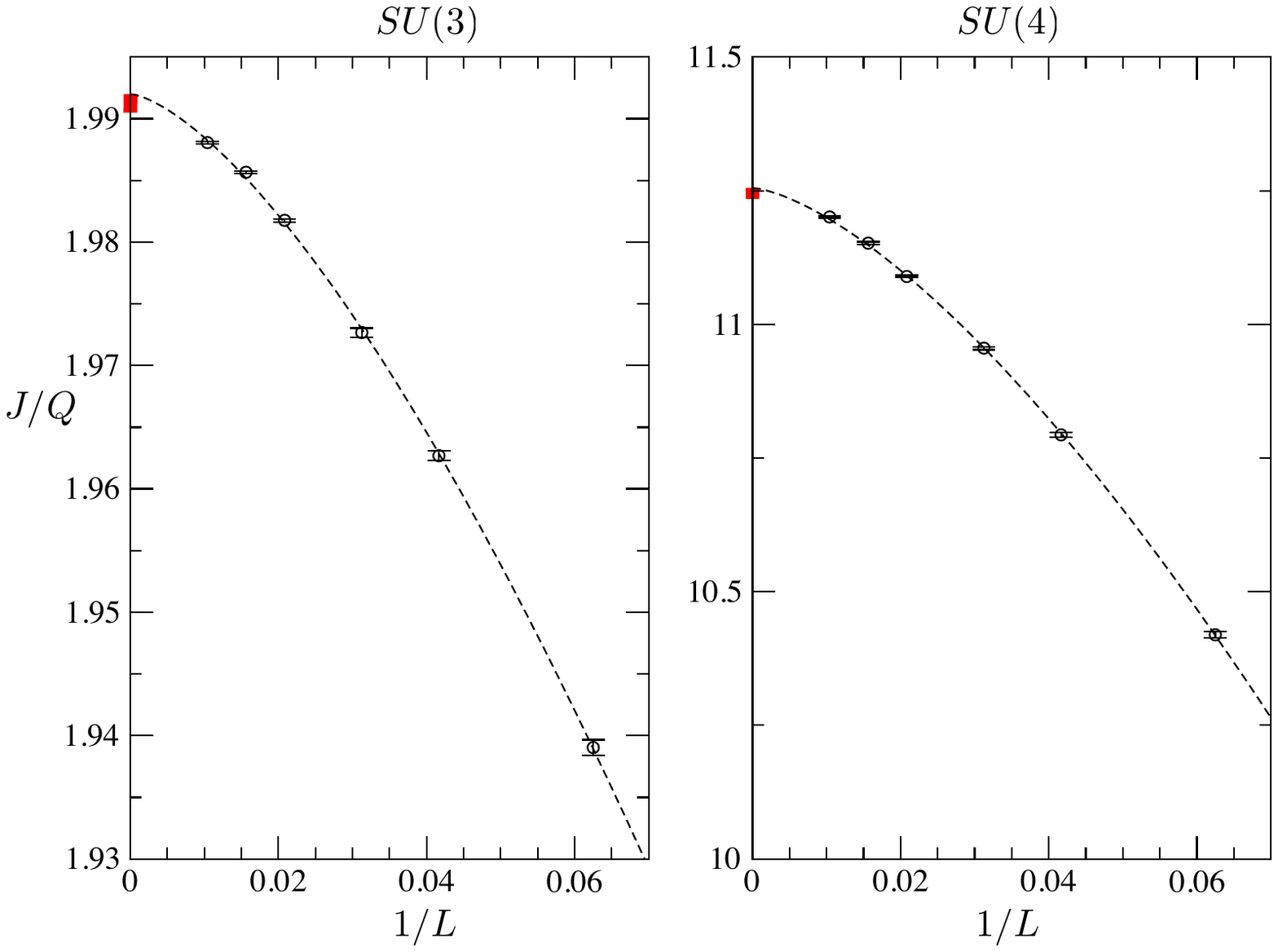}
    \caption{\label{fig:xint} The $x-$intercept ($J/Q$) of the crossing of $L$ and $2L$ for the stiffness shown in Fig.~\ref{fig:chirho_bare}, plotted versus $1/L$. The errors bars are estimated by a bootstrap method detailed in the text. The error bars get smaller for large systems because these curves are steeper; the $x-$intercept of the crossing can hence be determined more precisely. At $1/L=0$ the thick red line shows the bracket for the critical point from the analysis of the Binder ratio data (Fig.~\ref{fig:binder}). Clearly the $x-$intercept of the crossing converges to a finite value, and this value is completely consistent with the Binder ratio crossing. The dashed line is a non-linear fit to the form $g-A/L^a$, with $g=1.992, A=3.20,a=1.47$ for SU(3) and $g=11.255, A=51.3,a=1.48$. A motivation for this form is given in Appendix~\ref{sec:cross}.}
\end{figure}

\begin{figure}[t]
  \includegraphics[width=3.5in]{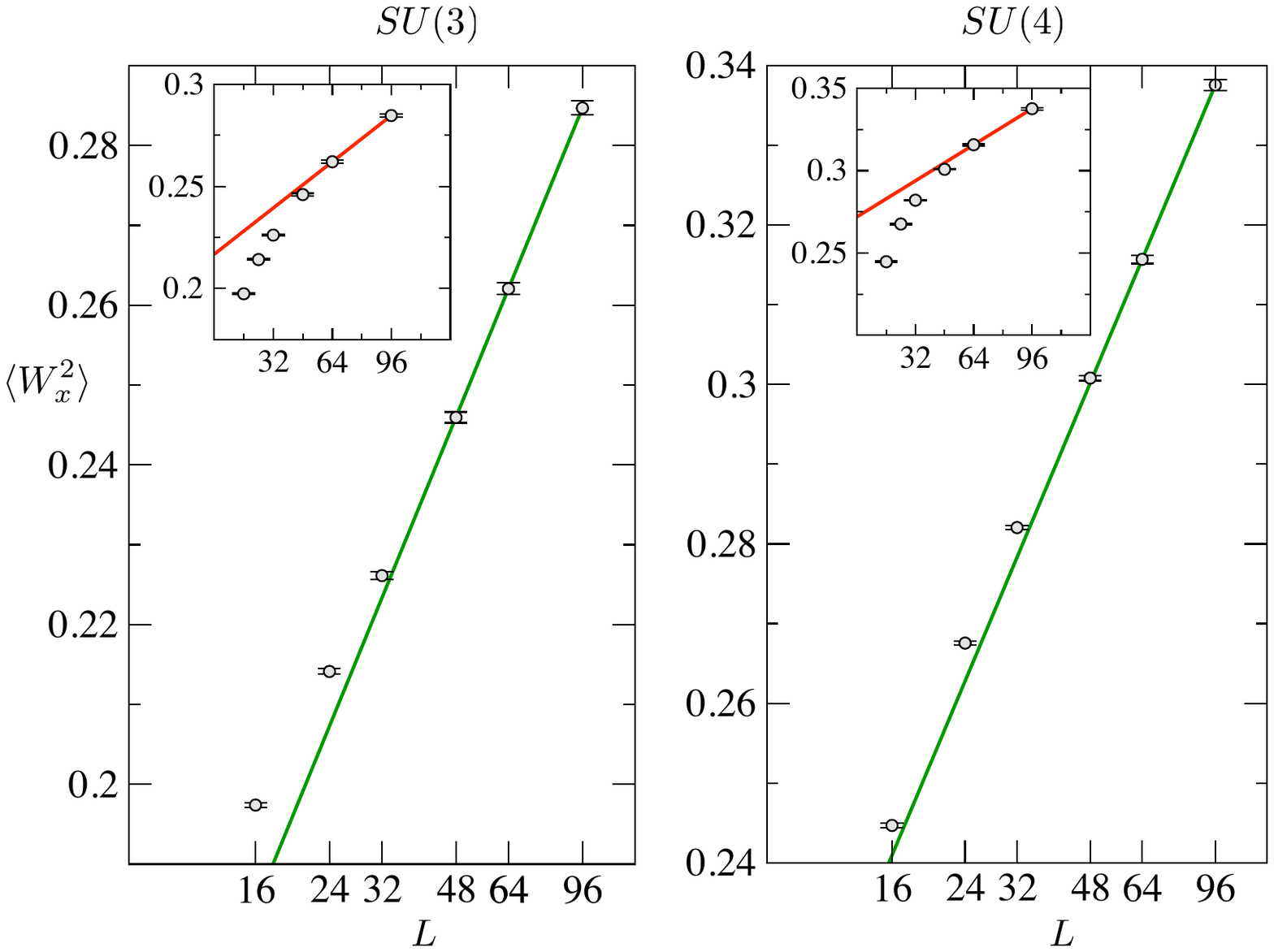}
    \caption{\label{fig:yint_loglin} Testing a linear and logarithmic divergences in the $y-$intercept of the crossing points. We have made the ``fits'' using {\em only} the two biggest system sizes (insets fit to $A+Bx$ and main panels $A+B{\rm ln}(x)$). This is thus more a test of the model than a fit, hence we refer to it as a ``fit''. The inset shows a linear ``fit'' (straight line on a lin-lin plot) and the main graphs a log ``fit'' (straight line on a log-lin plot). The insets clearly show that a linear divergence is completely inconsistent with our data. The main graphs show that a multiplicative log form fits the data reasonably, but is not as good as the power law (see Fig.~\ref{fig:yint_loglog})  }
\end{figure}

\begin{figure}[t]
  \includegraphics[width=3.5in]{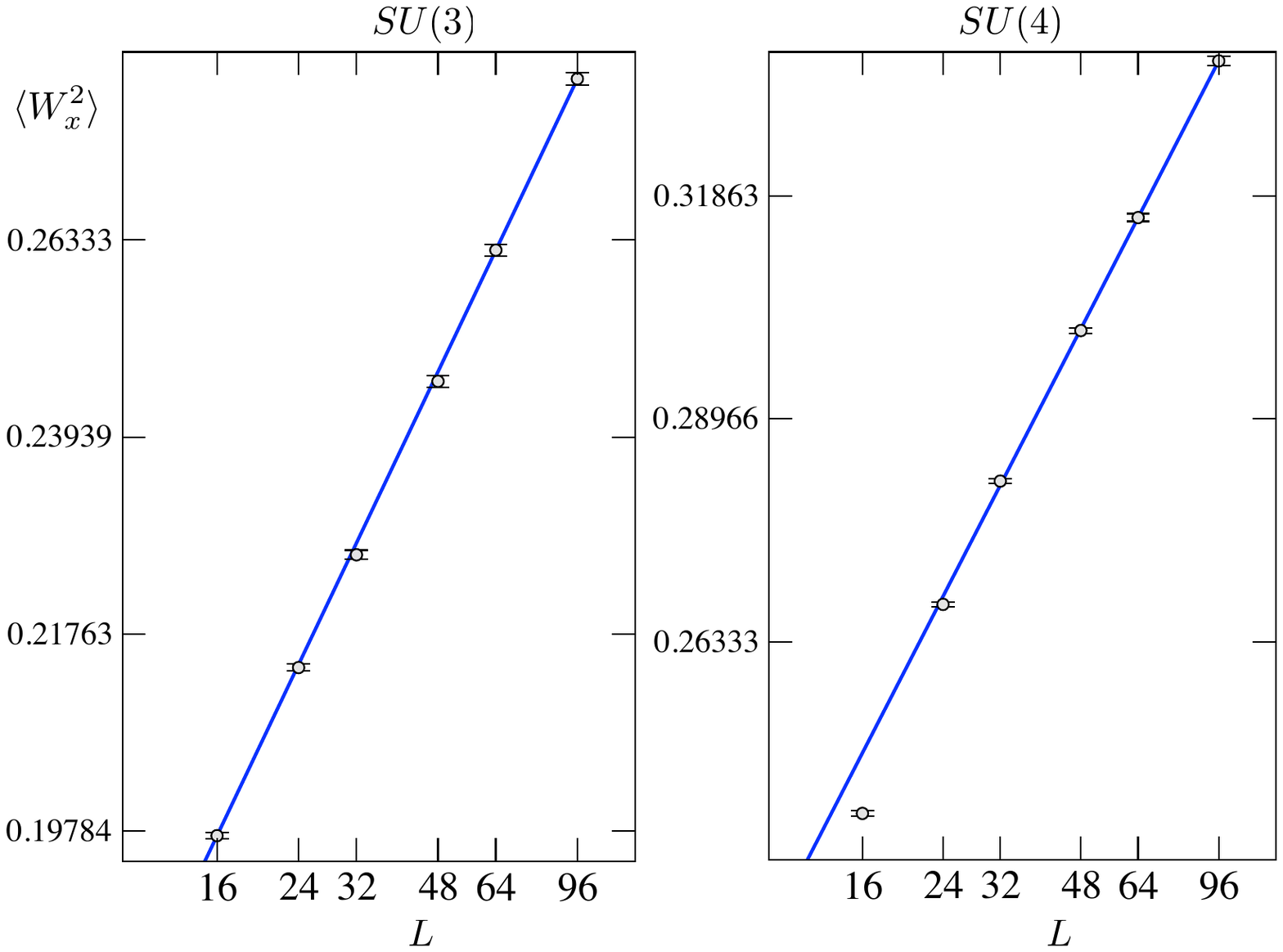}
    \caption{\label{fig:yint_loglog} Same as previous figure, but now a test of the possibility of a power law divergence. Again the straight line drawn is made by ``fitting'' the form $A x^B$ to the two largest $L$ points. This model fits the SU(3) data extremely well down to the smallest system size. For the SU(4) data the fit is also excellent, if one attributes the deviation of the smallest system size to a finite-size effect. The unfamiliar numbers on the $y-$axis are because we have chosen to tick and label the log axis with points on a base of $1.1$, which although convenient here, is not one of the familiar bases. Doing a standard linear regression analysis on the entire data set for SU(3) and the all the data except for the $L=16$ point for SU(4), we find estimates for the exponent, $B_{3}=0.20(2)$ and $B_{4}=0.16(2)$. The errors quoted here are based on roughly fitting the power law form to different data sets and making an estimate for how much the exponent differs. }
\end{figure}

Figs.~\ref{fig:chirho_bb} (on a bigger scale) and~\ref{fig:chirho_bare} (zoom-in close to the critical point) show $\langle W_x^2\rangle (g)$ for different $L$, for both the SU(3) and SU(4) model. While the data does cross, it does so very roughly. There are clearly large deviations from simple scaling behavior. Are these just corrections to scaling or is the leading scaling behavior itself affected? To make further progress in understanding the deviations from scaling, we study the crossing point of $\langle W_x^2\rangle$ for the sizes $L$ and $2 L$ in Fig.~\ref{fig:chirho_bare}. The crossing points have an $x-$intercept (the coupling $J/Q$) and a $y-$intercept (the value of $\langle W_x^2\rangle$). We study these intercepts as a function of $L$ with the hope of being able to extrapolate the large-$L$ behaviors. In order to estimate an error bar for these crossing points, we carry out a somewhat laborious bootstrap method, in which we generate synthetic data sets of the same length as our data (drawn by randomly sampling our own data), carry out least squares fit to this data (with polynomials of order two), and then compute the crossing point in each simulated data set. The RMS of the extracted crossing point data is then a reasonable estimate of the error in determining the $x-$ and $y-$intercepts of the crossing point. Fig.~\ref{fig:xint} shows the $J/Q$ value for the crossing points as a function of $1/L$. The data clearly converges for large $L$ to a finite value. It is re-assuring that this value is completely consistent with the brackets for the critical point extracted from the Binder ratio data (shown as thick lines on the $y-$axis of Fig.~\ref{fig:xint}). Next, we turn to the $y-$intercept of the crossings. Here things are much harder to interpret because the data does not saturate even for the largest system sizes studied here. Of course, one could always argue that this is due to a finite size effect, but given the large volumes simulated here, and the absence of any sign of saturation in the $y-$intercept, we shall assume that $\langle W_x^2\rangle$ diverges as the volume increases. We would like to note here that this by itself does not indicate a first order transition. At a first order transition $\langle W_x^2\rangle$ should grow linearly (because the stiffness itself remains finite instead of going to zero). We see in the insets of Fig.~\ref{fig:yint_loglin} that there is no evidence for a linear divergence. We find that for our data on the SU(3) and SU(4) models, a weak power law divergence seems to fit better (though only slightly better) than the logarithmic divergence found for SU(2)~\cite{sandvik2010:log}, see Figs.~\ref{fig:yint_loglin} and~\ref{fig:yint_loglog}. The power law divergence in the value of $\langle W_x^2\rangle$ ($y$-intercept) at the crossings and the simultaneous convergence of the value of $J/Q$ ($x$-intercept) at the crossing as the volume is increased can be understood in terms of multiplicative terms on the conventional scaling form for an object of zero scaling dimension. This argument is detailed in Appendix~\ref{sec:cross}.

\section{Discussion}
\label{sec:discuss}

We now turn to a summary of our interpretation of the numerical study presented in this paper.
\begin{itemize}
\item Just like in the SU(2) case for both SU(3) and SU(4) we find good evidence for a continuous transition and no evidence for first order behavior. This is in agreement with another study on this model~\cite{lou2009:sun}.

\item In contrast to a previous study~\cite{lou2009:sun} in the SU(3) and SU(4) models which have used conventional scaling forms for SU(3), we find evidence for {\em multiplicative} terms to the conventional scaling form for SU(3) and SU(4) that affect the leading behavior, qualitatively similar to those found in the SU(2) case~\cite{sandvik2010:logs}. A speculative discussion on whether these deviations are due to multiplicative weak power laws or whether they are multiplicative logarithms is given below -- It is difficult to distinguish unambiguously between these two possibilities solely based on numerics. 

\item The stiffness (and susceptibility) have large deviations from simple scaling. The crossing point converges on the $x-$axis to the same critical coupling deduced from the Binder ratio, whereas the value of the $L\rho_s$ itself ($y-$axis) has a sizable volume dependence. If we assume the observed behavior
in Fig.~\ref{fig:yint_loglin} is the leading scaling behavior, we conclude that the winding number fluctuations diverges sub-linearly. The fact that the divergence is sub-linear is very crucial, since it implies that the stiffness (which is the winding number fluctuation divided by $L$) itself goes to zero as one approaches the critical point, although not as $1/L$
like one would have expected from a naive scaling analysis. 

\item We have also argued that a scaling from with a {\em multiplicative} power law term can explain both the salient features of the $\langle W_x^2\rangle$ crossings, {\em i.e.}, first that the $x$-intercept ($J/Q$) of the crossing converges to the critical coupling as the volume is increased and second that the $y-$intercept ($\langle W_x^2\rangle$) has a power law divergence.
\end{itemize}

Perhaps, the simplest explanation for the stiffness measurement is that the volume dependence of the stiffness curve is a result of an {\em additive power law} correction from an irrelevant operator and that the stiffness curve would saturate as the system size is made even larger than that studied here -- this would be the expectation from a conventional scaling hypothesis for a critical point. Our numerics strongly suggest that this scenario is unlikely given the large system sizes studied here and the absence of any indication of saturation in the stiffness. 

We are thus forced to consider {\em multiplicative} deviations from simple scaling: What could result in this unusual scaling for the stiffness? This behavior must signal the breakdown of one of the standard scaling assumptions made.   One scenario which would result in the divergence of $L\rho_s$ is if the scaling function depended non-analytically on an irrelevant operator. For instance, imagine that $L\rho_s= f(gL^{1/\nu}, \frac{g_\omega}{L^\omega} )$ with $\omega>0$ and hence $g_\omega$ formally irrelevant.
Normally one assumes the scaling function is analytic in its arguments and can hence set the second argument to zero. If it was not analytic and depended for instance like $f(x,y)\approx \frac{1}{y}$ for small $y$, there would be a power law divergence in $L\rho_s$ as observed here. One important consequence of this scenario is that a power law divergence is 
more natural than a log divergence, since  a log divergence (within this scenario) would require a singular dependence of the scaling function and an exactly marginal operator in addition.
We leave it for future work to strengthen this speculative argument, by
understanding the nature of the irrelevant operator and the reasons it results in singular scaling functions for the stiffness.

An interesting discussion is how the critical point evolves from the SU(2), SU(3) to SU(4) critical points. Since the first theoretical work about such critical points~\cite{senthil2004:science}, it has been assumed that the properties of the critical point depend smoothly on $N$. As stated in the beginning of this section, we find that this is indeed the case qualitatively. Anomalous deviations that affect the leading behavior of the stiffness were found in studies of the SU(2) model. In that study it was concluded that the leading divergence is governed by a {\em multiplicative logarithm}, here we propose that the leading divergence is governed by a {\em multiplicative weak power law}. Only using numerical techniques it is not possible to distinguish between these two without access to decades of data (which we do not have at the current time). 
If we assume that the behavior found for SU(2)~\cite{sandvik2010:logs} results from a marginal operator, it is natural to assume that the scaling dimension of this operator should depend on $N$ and it would hence become irrelevant at larger $N$. According to the non-analytic scaling function argument made above, this would make it natural to have a power law divergence at the SU(3) or SU(4) critical points. It is also possible that in the SU(2) case a weak power law divergence of the same form as proposed here for SU(3) and SU(4) is present. A complete theoretical scenario is required to address these 
interesting possibilities.

{\em Note added:} During the review for publication of this work, a study of the $T=0$ data on impurity scaling in the SU(3) with the valence bond projector QMC method has appeared~\cite{banerjee2010:su3} . We are both in agreement that the SU(3) model has a continuous transition.  The authors have however analyzed the impurity data based on a conventional scaling hypothesis which has only {\em additive} scaling corrections: they have not found the {\em multiplicative} scaling corrections in the impurity problem that we have found here in the bulk. Whether this a fundamental disagreement or a result of studying different ({\em i.e.}  impurity versus bulk) problems is not clear to the author.

\section{Acknowledgments}

The author is grateful to Roger Melko for collaborations on related work and to Anders Sandvik for a number of useful discussions.
He has also benefited from discussions with L. Balents, M. Fisher, N. Kawashima, H. Katsura, M. Levin, J. Lou, O. Motrunich,  and T. Senthil. 
All the simulation reported here were
carried out on the BCX and DLX clusters at the Center for Computational Sciences at the University of Kentucky. The author acknowledges support from NSF DMR-1056536.

\appendix

\section{Method and algorithm}
\label{sec:method}
The method employed here is the stochastic series expansion, which is well documented in the literature~\cite{syljuasen2002:dirloop}.
In the way we have defined the Hilbert space, one can think of the configuration space as a set of loops (the loops travel vertically up in time on the A sub-lattice and vertically down in time on the B sub-lattice) with a color assignment from one of $N$ colors.  
For the updates, we have generalized the 
``deterministic'' method of Ref.~\onlinecite{syljuasen2002:dirloop}, Sec. II D, to both the four site Q term and SU(N), this is possible since as noted in Sec.~\ref{sec:model}, because of the simple form of the matrix elements we are guaranteed that all configurations which appear at a given order of $J$ and of $Q$, appear with the same weight. 
In this method there are two kinds of updates (1) [loop update] the $N$ colors of a loop can 
be randomly assigned a new color  and (2) [diagonal update] swap of identity operators with diagonal operators.

\begin{figure}[!t]
\includegraphics[width=3.5in]{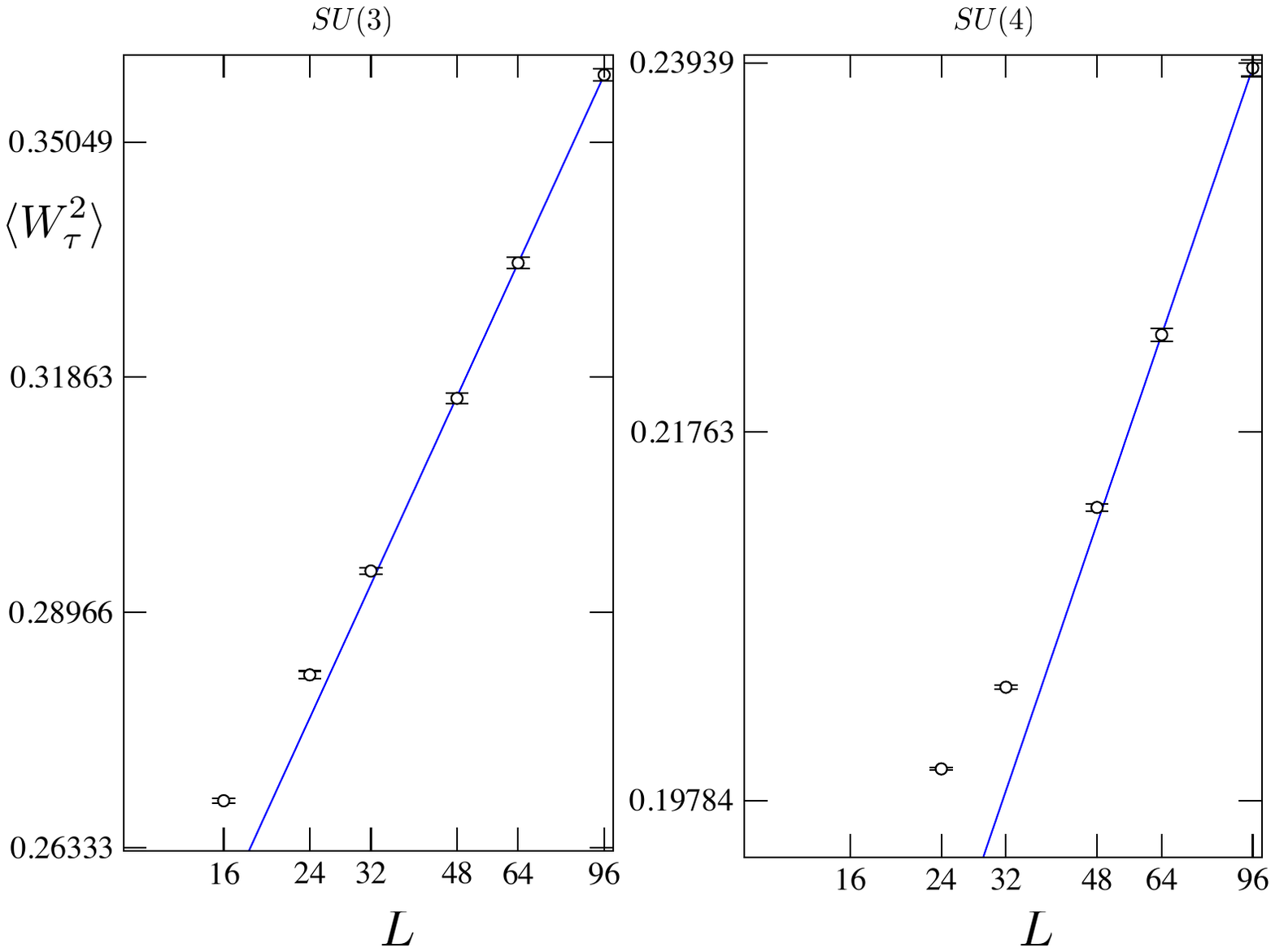}
  \caption{\label{fig:yint_susc} An identical analysis as that leading up to Fig.~\ref{fig:yint_loglog}, but now for the fluctuations of the temporal winding number. The power law ``fit'' does not work quite as well here. Again the ``fit'' here is carried out by fitting the power law form $Ax^B$ to the two largest system sizes. The exponents we find here are $B_3=0.18(2)$ and $B_4=0.17(2)$, where the errors are determined roughly by fitting different sets of the data. Reassuringly, they are in agreement with the values found for the stiffness. We note here that a fit to a multiplicative logarithm form (not shown) for $\langle W_\tau^2\rangle$ does just as well. See also Fig.~\ref{fig:ratio} for a comparison between the susceptibility and stiffness. }
\end{figure}

\section{Choice of Cartan algebra}
\label{sec:Cartan}

In this appendix we list our choices for the $N-1$ diagonal generators for $N=3,4$. There are different conventions with which these can be chosen.
For SU(2) there is only one diagonal generator and the natural choice is:
\[
\begin{pmatrix}
1/2 & 0  \\
0 & -1/2
\end{pmatrix} \]

for SU(3) we used:
\[\begin{pmatrix}
1/2 & 0 & 0 \\
0 & -1/2 & 0 \\
0 & 0 & 0
\end{pmatrix} \]

\[\begin{pmatrix}
1/(2\sqrt{3}) & 0 & 0 \\
0 & 1/(2\sqrt{3}) & 0 \\
0 & 0 & -2/(2\sqrt{3})
\end{pmatrix} \]

Finally for SU(4) we used:
\[\begin{pmatrix}
1/(2\sqrt{2}) & 0 & 0 &0\\
0 & -1/(2\sqrt{2}) & 0 &0 \\
0 & 0 & 1/(2\sqrt{2}) & 0\\
0 &0 & 0 & -1/(2\sqrt{2}) 
\end{pmatrix} \]

\[\begin{pmatrix}
-1/(2\sqrt{2}) & 0 & 0 &0\\
0 & 1/(2\sqrt{2}) & 0 &0 \\
0 & 0 & 1/(2\sqrt{2}) & 0\\
0 &0 & 0 & -1/(2\sqrt{2}) 
\end{pmatrix} \]

\[\begin{pmatrix}
-1/(2\sqrt{2}) & 0 & 0 &0\\
0 & -1/(2\sqrt{2}) & 0 &0 \\
0 & 0 & 1/(2\sqrt{2}) & 0\\
0 &0 & 0 & 1/(2\sqrt{2}) 
\end{pmatrix} \]

All these choices are made so that they satisfy the standard normalization,
 ${\rm Tr}[X^aX^b]=\delta_{ab}/2$.\\

\section{Ratio of size and temperature}
\label{sec:lt}

Assuming $z=1$ scaling, a choice has to be made for the ratio of $LT/c$.
Of course any fixed ratio will work for the scaling properties, but in order to
best use the efforts of our simulations we would like to work as close the cubic limit as possible $LT/c=1$.
The problem is that we 
do not know what $c$ is in terms of our couplings $J$ and $Q$, {\em a priori}. To circumvent this problem, we pick the temperature so that the 
fluctuations of the winding number in space (the re-scaled spin stiffness) and the fluctuations in the winding number in time
(the rescaled spin susceptibility) are of the same order of magnitude. Indeed we would expect them to be identical at the isotropic point. 
This is achieved most simply by choosing a coupling close to the critical
point and the choosing some appropriate system size (small is good enough, so that the correlation length is larger than the system size)  
and then adjusting the temperature till the fluctuations of the temporal and spatial winding number become approximately equal. We found
that working in units with  $Q=1$ that the approximate equality of the temporal and spatial winding numbers was achieved for $ L/\beta=4$ for SU(3) and for $ L/\beta=10$ for the SU(4) case, we hence used these values for the simulations reported in this manuscript.

\begin{figure}[!t]
\includegraphics[width=3.5in]{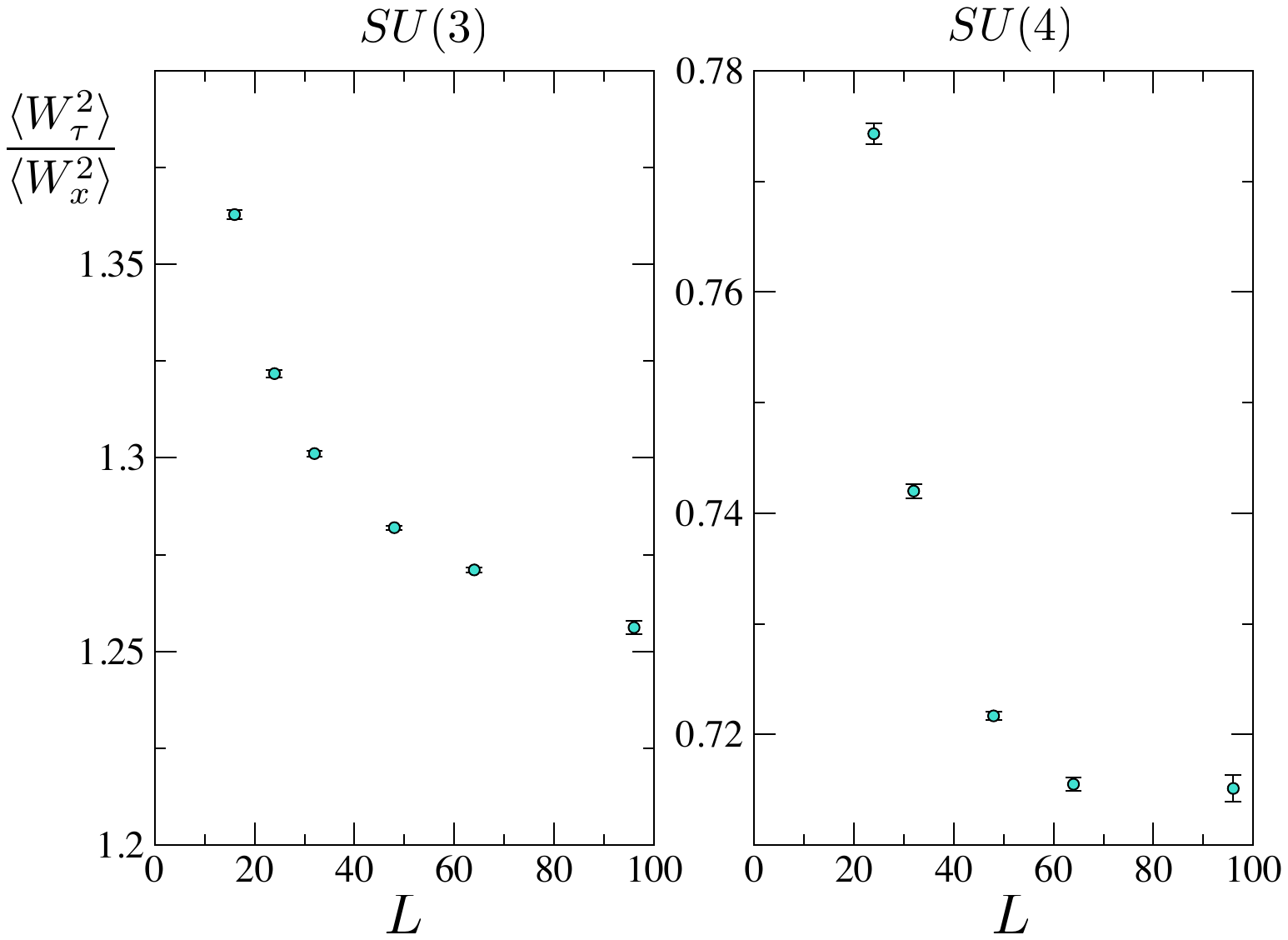}
 \caption{\label{fig:ratio} The ratio of the fluctuations of the temporal and spatial winding numbers  evaluated at the crossing points of the curves for the susceptibility. From the saturation to a {\em finite} (non-zero, non-infinite) value of the ratio at large volume shown here, it is clear that in the large volume limit the asymptotic behavior is identical for $\langle W_x^2\rangle$ and $\langle W_\tau^2\rangle$. The ratio is not expected to go exactly to $1$ away from the $LT/c=1$ point. The number is of order one, because of our choice of $LT$ as discussed in Appendix~\ref{sec:lt}.  }
\end{figure}

\section{Crossing point}
\label{sec:cross}

In this appendix we study how the crossing analysis is affected by {\em a multiplicative correction} to the standard scaling function of an object with zero scaling dimension. Our main result here is that multiplicative scaling corrections do not affect the fact that the x-value of the crossing converges to the critical point in the thermodynamic limit but they affect the leading behavior of the y-value of the crossing, which does not saturate as one would expect in the absence of multiplicative power laws.

Imagine we had a scaling function that had the form, $L^\alpha f(g L^a)$. Now lets ask for what value of $g_x$ it crosses.
Assuming the function $f$ is analytic in its argument we find close to the critical point,
 $g_x=\frac{f(0)}{L^a f^\prime(0)}
\frac{\Lambda^\alpha-1}{1-\Lambda^{a+\alpha}}
$ for the crossing of curves of sizes $L$ and $\Lambda L$. So, the value of the
critical coupling does go to zero (the true critical point) for larger and larger $L$, but it already receives corrections
without any corrections from irrelevant operators. The leading behavior of the quantity at the crossing point, described by such a scaling function is 
 simply, $L^\alpha f(0)$. In order to test the functional form of $g_x$ we have carried out a non-linear fit to the data on the stiffness as detailed
in the caption of Fig.~\ref{fig:xint}. We find $\nu\approx0.65(5)$ for both SU(3) and SU(4) which is consistent with the value of $\nu$ reported in a previous study~\cite{lou2009:sun} (the error quoted here for $\nu$ has been determined very roughly by attempting a number of non-linear fits dropping
different sets of data points). Interestingly the $\nu$ in the previous work~\cite{lou2009:sun} was extracted
completely from the data on the order parameters. This gives some credibility to the idea that there is a multiplicative correction to the naive scaling for the stiffness.
The analysis is very similar for $\ln (L) f(gL^a)$, where again $g_x\approx \frac{f(0)}{\ln(L) L^a f^\prime(0)}$ for large $L$ and the function 
would be described by, $\ln(L)f(0)$ at leading behavior. Our $g_x$ data would fit this model reasonably too, since the multiplication of a log affects the numerical values very weakly.

\section{Susceptibility}

In this appendix we study the fluctuations of the temporal winding number (the re-scaled susceptibility). We carry out an identical analysis to that presented in the body of the text for $\langle W_x^2\rangle$.

The susceptibility has an identical crossing behavior to that of the stiffness shown in Fig.~\ref{fig:chirho_bare}; there is a drift in the crossing point in both the $x-$ and $y-$intercepts. We have carried out an identical analysis of the crossing of $L$ and $2L$ and we find that the $x-$intercepts (shown in Fig.~\ref{fig:xint} for the stiffness), do not depend very much on whether we look at the crossing of the susceptibility or the stiffness. The $y-$intercept does however, so we have made a separate plot for this quantity in Fig.~\ref{fig:yint_susc}. We find that unlike in the case of stiffness there are finite-size correction to the power law model. Yet the power of the fitted power law for the largest system sizes is in agreement with our findings for the susceptibility, as detailed in the inset. Finally, as an independent check that these two quantities diverge the same way as we approach the critical point, we have plotted the ratio of these two quantities at the crossing point value determined by the susceptibility crossing. This is shown in Fig.~\ref{fig:ratio}. As expected we find that the ratio of the quantities clearly goes to a constant at large volume, providing clear evidence that the two quantities diverge the same way, as one would expect for the emergent $z=1$ scaling (we use the word emergent here, since the microscopic model is clearly not invariant under space-time rotations!).

\bibliography{/Users/ribhukaul/OPPIE/Physics/PAPERS/BIB/career}

\end{document}